\documentclass[seceq]{ptptex}

\usepackage{graphicx}

\newcommand{\arXivid}{arXiv:}

\newcommand{\hepph}{hepph/}
\newcommand{\astroph}{astroph/}
\newcommand{\beq}{\begin{equation}}
\newcommand{\eeq}{\end{equation}}
\newcommand{\baq}{\begin{eqnarray}}
\newcommand{\eaq}{\end{eqnarray}}
\newcommand{\fnl}{f_{\mathrm{NL}}}
\newcommand{\gnl}{g_{\mathrm{NL}}}
\newcommand{\rdec}{r_{\mathrm{dec}}}
\newcommand{\rhor}{\rho_\mathrm{r}}
\newcommand{\Mp}{M_\mathrm{Pl}}
\newcommand{\bm}{\bf }
\title{
The self-interacting curvaton%
}

\author{
Kari \textsc{Enqvist}%
}

\inst{
Physics Department, University of Helsinki, and Helsinki Institute of Physics,
PO. Box 64, FIN-0014 University of Helsinki, Finland
}

\abst{
The evolution of the curvature perturbation is highly non-trivial
for curvaton models with self-interactions and is very sensitive to
the parameter values. The final perturbation depends also on the
curvaton decay rate $\Gamma$. As a consequence, non-gaussianities
can be greatly different from the purely quadratic case, even if the
deviation is very small. Here we consider a class of
polynomial curvaton potentials and discuss the dynamical behavior of
the curvature perturbation. We point out that, for example, it is
possible that the non-gaussianity parameter $\fnl\simeq 0$ while $\gnl$ is non-zero. In the case
of a curvaton with mass $m\sim {\cal O}(1)$ TeV we show that one
cannot ignore non-quadratic terms in the potential, and that only a
self-interaction of the type $V_{\rm int}=\sigma^8/M^4$ is
consistent with various theoretical and observational constraints.
Moreover, the curvaton decay rate should then be in the range
$\Gamma=10^{-15}- 10^{-17}$ GeV. }

\begin{document}

\maketitle

\section{Introduction}
In the curvaton mechanism \cite{curvaton}, primordial
perturbations originate from quantum fluctuations of a light scalar
field which gives a negligible contribution to the total energy
density during inflation. This field is called the curvaton
$\sigma$. Inflation is driven by another scalar, the inflaton
$\phi$, whose potential energy dominates the universe. After the end
of inflation, the inflaton decays into radiation. If the
inflationary scale is low enough, $H_{*}\ll 10^{-5}
\sqrt{\epsilon_{*}}$, the density fluctuations of the radiation
component are much below the observed amplitude $\delta\rho/\rho\sim
10^{-5}$ and the fluid is for practical purposes homogeneous. While
the dominant radiation energy scales away as $\rho_{r}\sim a^{-4}$,
the curvaton contribution to the
total energy density may increase and the initially
negligible curvaton perturbations get imprinted into the metric. The standard adiabatic hot big bang era is recovered
when the curvaton eventually decays and thermalizes with the
existing radiation. The mechanism can be seen as a conversion of
initial isocurvature perturbations into adiabatic ones and,
depending on the parameters of the model, is capable of generating
all of the observed primordial perturbation.
The scenario sketched above represents the simplest possible
realization of the curvaton mechanism, and a wide range of different
variations of the idea have been studied in the literature. For
example, the inflaton perturbations need not be negligible
\cite{mixed}, there could be several curvatons
\cite{many_curvatons}, the curvaton decay can result into residual
isocurvature perturbations \cite{LUW,isocurvature} and inflation
could be driven by some other mechanism than slowly rolling scalars
\cite{alt_inflation}.

It is however well known that the
predictions of the curvaton model are quite sensitive to the form of
the curvaton potential \cite{Dimopoulos:2003ss,kesn,kett,Enqvist:2008be,Huang:2008zj,Kawasaki:2008mc,Chingangbam:2009xi,us1,us2,Chambers:2009ki, us3}.
In particular, even small deviations from
the extensively studied quadratic potential can have a significant
effect, at least when considering non-Gaussian effects \cite{kesn,
kett}. One can also encounter strong scale-dependence of the non-gaussianity parameters \cite{byrnesetal}.
When the initial curvaton field value lies far in
the non-quadratic part of the potential, the non-linear nature of
the evolution equation will in general result in a very rich
structure of phenomena in the parameter space, as has been
discussed in detail in \cite{us1,us2}.

The simplest non-quadratic curvaton potential is given by
\begin{equation}
V = \frac{1}{2}m^2\sigma^2 + \lambda{\sigma^{n+4}\over M^n} \ ,
\label{curvatonpot}
\end{equation}
where $n$ is an even integer to keep the potential bounded from
below, and the interaction term is suppressed by a cut-off scale $M$.
For non-renormalizable operators $n > 0$, we set the cut-off scale to
be the Planck scale $M = M_{\rm P}\equiv 1$, and the coupling to
unity, $\lambda = 1$. For the  renormalizable quartic case $n = 0$,
the coupling $\lambda$ can be treated as a free parameter.

The potential
(\ref{curvatonpot}) is reasonably well motivated by generic theoretical
arguments. Indeed, the curvaton should have interactions of some
kind as it eventually must decay and produce Standard Model fields.
The curvaton needs to be weakly interacting to keep the field light
during inflation. This however only implies that the effective
curvaton potential should be sufficiently flat in the vicinity of
the field expectation value during inflation but does not a priori
require the interaction terms in (\ref{curvatonpot}) to be
negligible. Moreover, as typically the inflationary energy scale is relatively
high, the field can be displaced far from the origin and therefore
feels the presence of the higher order terms in the potential. The
interactions could arise either as pure curvaton self-interactions
involving the curvaton field $\sigma$ alone, or more generically as effective
terms due to curvaton couplings to other (heavy) degrees of freedom
that have been integrated out. An example of a possible physical
setup which could lead to (\ref{curvatonpot}) is given by flat
directions of supersymmetric models that have been suggested as
curvaton candidates \cite{curvaton_flat}. These would lead to a
potential of the form (\ref{curvatonpot}) with typically a
relatively large power for the non-renormalizable operator.

The amplitude and non-gaussianity of the perturbation depend
on the curvaton decay time. Here we
assume for simplicity a perturbative curvaton decay
characterized by some effective decay width $\Gamma$ (for non-perturbative decay, see \cite{curvatondecres, BasteroGil:2003tj}).

When the interaction term dominates in
(\ref{curvatonpot}), the curvaton oscillations start in a
non-quadratic potential and the curvaton energy density always
decreases faster than for a quadratic case. For non-renormalizable
interactions, the decrease is even faster than the red-shifting of
the background radiation and the curvaton contribution to the total
energy density is decreasing at the beginning of oscillations.
Consequently, the amplification of the curvaton component is less
efficient than for a quadratic model. For the same values for $m$
and $\Gamma$, the curvaton typically ends up being more subdominant
at the time of its decay than in the quadratic case.

Despite the subdominance, the curvaton scenario can yield the
correct amplitude of primordial perturbations as the relative
curvaton perturbation $\delta\sigma_{*}/\sigma_{*}$ produced during
inflation can be much larger than $10^{-5}$. For a quadratic model,
it is well known that the curvaton should make up at least few per
cents of the total energy density at the time of its decay in order not
to generate too large non-gaussianities \cite{LUW,curvaton_ng}. This
bound does not directly apply to the non-quadratic model
(\ref{curvatonpot}) since the dynamics is much more complicated.
Although the subdominant curvaton scenario implies relatively large
perturbation $\delta\sigma_{*}/\sigma_{*}$, the higher order terms
in the perturbative expansion of curvature perturbation can be
accidentally suppressed \cite{kesn,kett, us2}.

\section{The curvature perturbation}
\label{nongausscurv}

Let us adopt the $\delta N$ formalism \cite{deltaN,recent_deltaN} and assume
that the curvature perturbation $\zeta$ arises solely from the inflation generated perturbation of a
single curvaton field. Then
\beq
\label{zeta}
\zeta(t,{\bm x}) = N'(t,t_{*})\delta\sigma_{*}({\bm x})
+\frac{1}{2}N''(t,t_{*})\delta\sigma_{*}({\bm x})^2
+\frac{1}{6}N'''(t,t_{*})\delta\sigma_{*}({\bm x})^3 \cdots \, .
\eeq
Here $N(t,t_{*})$ is the number of e-foldings from an initial
spatially flat hypersurface with fixed scale factor $a(t_*)$ to a
final hypersurface with fixed energy density $\rho(t)$, evaluated
using the FRW background equations. The final time $t$ is some arbitrary time after the curvaton decay. The prime denotes a
derivative with respect to the initial curvaton value
$\sigma_{*}$. Here we take $t_{*}$ to be some time during inflation
soon after all the cosmologically relevant modes have exited the
horizon and assume that the curvaton perturbations $\delta\sigma_{*}$ are
Gaussian at this time. The expansion (\ref{zeta}) is then of the form
\beq
 \label{fnl_gnl}
 \zeta(t,{\bm x})
 =
 \zeta_{\rm g}(t,{\bm x})
 +\frac{3}{5}f_{\rm NL}\zeta_{\rm g}(t,{\bm x})^2
 +\frac{9}{25}g_{\rm NL}\zeta_{\rm g}(t,{\bm x})^3+\cdots \ .
\eeq
where $\zeta_{\rm g}(t,{\bm x})$ is a Gaussian field and the
non-linearity parameters are given by
\begin{eqnarray}
\label{fnl_def}
\fnl &=& \frac{5}{6}\frac{N''}{N'^2}~,\\
\label{gnl_def}
\gnl &=& \frac{25}{54}\frac{N'''}{N'^3}\ .
\end{eqnarray}
Here we neglect all the scale dependence of the non-linearity parameters \cite{Byrnes:2009pe, byrnesetal}. With this assumption, and neglecting higher order perturbative corrections, 
the constants $\fnl$ and $\gnl$ measure the amplitudes of the three- and four-point correlators of $\zeta$, respectively. Observationally,
they can be extracted from the CMB bi- and trispectra.

We assume the curvaton obeys slow roll dynamics during inflation and
introduce a parameter $r_{*}$ to measure its contribution to the total
energy density $\rho$ at $t_{*}$:
\beq
 \label{r_star}
 r_{*}=\frac{\rho_{\sigma}}{\rho}\Big|_{t_{*}}\simeq\frac{V(\sigma_{*})}{3H_{*}^2}\ll 1\ .
\eeq
Here the inflationary scale $H_{*}$ is a free parameter, up to certain model dependent consistency
conditions. Assuming inflation is driven by a slowly rolling inflaton
field, we need to require $H_{*}\ll 10^{-5} \sqrt{\epsilon}$ in order
to make the inflaton contribution to $\zeta$ negligible. In this setup
we also need to adjust the slow-roll parameter $\epsilon=-\dot{H}_{*}/H_{*}^2$, determined by
the inflaton dynamics, to give the correct spectral index\cite{Wands:2002bn},
$n-1=2\epsilon-2\eta_{\sigma\sigma}$. The curvaton
contribution, $\eta_{\sigma\sigma}=V''(\sigma_{*})/3H_{*}^2$, is
typically negligible because of the subdominance of the curvaton. The curvaton mass is required to be small, but the same also holds for the inflaton mass.

After the end of inflation, one assumes that the inflaton decays completely
into radiation and the universe becomes radiation dominated. The decay constant $\Gamma$ accounts for
the coupling between the radiation and the curvaton component. The
evolution of the coupled system is then given by
\begin{eqnarray}
\label{frw1}
&&
\ddot{\sigma} + (3H+\Gamma)\dot{\sigma}+m^2\sigma + \lambda(n+4)\sigma^{n+3} = 0\\
&&
\dot{\rho_\mathrm{r}} = -4H\rho_\mathrm{r} + \Gamma \dot{\sigma}^2\\
&&
3H^2 = \rho_\mathrm{r} + \rho_\sigma \ .
\end{eqnarray}
The initial conditions are given by $\rho_{\mathrm{r}} = 3H_*^2$ and
$\rho_\sigma = V(\sigma_*) = r_* / (1-r_*) \rho_{\mathrm{r}}$ specified at time $t_*$ corresponding to the end of inflation. We also set
$\dot{\sigma} = 0$. Given the parameters $n$, $\lambda$ and $m$, which
determine the potential (\ref{curvatonpot}), and the two initial conditions
$r_*$ and $H_*$, one can calculate $N$ in (\ref{zeta}) from this set of
equations. To find the curvature perturbation, we set
$\delta\sigma_{*}=H_{*}/(2\pi)$ and compute
$\zeta=N(\sigma_{*}+\delta\sigma_{*})-N(\sigma_{*})$. For a given set
of parameters, one then adjusts the decay width $\Gamma$ so that the observed
amplitude\cite{wmap} $\zeta \sim 10^{-5}$ is obtained.

One may treat $\Gamma$ as a free parameter since we have not specified
the curvaton couplings to other matter, in particular to the Standard
Model fields. However, since the primordial perturbation is mostly adiabatic, the curvaton
should decay before dark matter decouples in order not to produce
isocurvature modes. Assuming a freeze-out temperature $T\simeq m_{\rm DM}/20$ for
an LSP type dark matter model with the LSP mass
$m_{\rm DM}\sim~{\cal O}(100)$ GeV, this translates to a rough bound
\begin{equation}
\label{eq:isocurvature}
\Gamma \gtrsim 10^{-17} \mathrm{GeV}\ .
\end{equation}
While this bound could be relaxed in non-minimal constructions, let us here
adopt (\ref{eq:isocurvature}) for definiteness.

\section{Small deviations from the quadratic form}
\label{smalldev}

Let us first assume that the deviation from
the quadratic from is small and write the curvaton potential (\ref{curvatonpot}) as\cite{kesn,kett}
\begin{equation}
\label{eq:V}
V(\sigma)
=
\frac{1}{2} m^2 \sigma^2
+
\lambda m^4 \left( \frac{\sigma}{m} \right)^n~,
\end{equation}
where $\lambda$ is some coupling constant.  It is also useful to define a parameter $s$ which
represents the size of the non-quadratic term relative to the
quadratic one:
\begin{equation}
\label{eq:def_s}
s \equiv 2 \lambda \left( \frac{\sigma_\ast}{m} \right)^{n-2}.
\end{equation}
Thus the larger $s$ is, the larger is the contribution from the
non-quadratic term.

The curvature fluctuation
can then be written, up to the third order, as \cite{Sasaki:2006kq}
\begin{eqnarray}
\zeta = \delta  N &=&
\frac{2}{3} r \frac{\sigma'_{\rm osc}}{\sigma_{\rm osc}}  \delta \sigma_\ast
+
\frac{1}{9} \left[ 3r\left(
1 +
\frac{\sigma_{\rm osc} \sigma_{\rm osc}^{\prime\prime}}{\sigma_{\rm osc}^{\prime 2}}
\right)
- 4 r^2 -2  r^3
\right]
\left( \frac{\sigma'_{\rm osc}}{\sigma_{\rm osc}} \right)^2  (\delta \sigma_\ast )^2 \notag \\
&&+
\frac{4}{81} \left[
\frac{9r}{4}  \left(
\frac{\sigma_{\rm osc}^2 \sigma_{\rm osc}^{\prime\prime\prime}}
{\sigma_{\rm osc}^{\prime 3}}
+
3\frac{\sigma_{\rm osc} \sigma_{\rm osc}^{\prime\prime}}{\sigma_{\rm osc}^{\prime 2}}
\right)
-9r^2
\left(
1
+
\frac{\sigma_{\rm osc} \sigma_{\rm osc}^{\prime\prime}}{\sigma_{\rm osc}^{\prime 2}}
\right)
\right. \notag \\
&&
\left.
+\frac{r^3}{2} \left(
1
-
9\frac{\sigma_{\rm osc} \sigma_{\rm osc}^{\prime\prime}}{\sigma_{\rm osc}^{\prime 2}}
\right)
+10r^4 + 3r^5
\right]
\left( \frac{\sigma'_{\rm osc}}{\sigma_{\rm osc}} \right)^3  (\delta \sigma_\ast )^3~,
\end{eqnarray}
where $\sigma_{\rm osc}$ is the value of the curvaton at the onset of
its oscillation; above\footnote{Note that in the literature there is much variation in the definition of $r$,
as is also in the present paper. }
\begin{equation}
\label{eq:def_r}
\left.
r \equiv \frac{3 \rho_\sigma}{4 \rho_{\rm rad} + 3\rho_\sigma}\right|_{\rm decay}.
\end{equation}
Notice that $\sigma^\prime_{\rm osc}/ \sigma_{\rm osc}=1/\sigma_\ast$
for the case of the quadratic potential.  With this expression, we can
write down the non-linearity parameter $f_{\rm NL}$ as
\begin{equation}
\label{eq:fNL}
f_{\rm NL} = \frac{5}{4r} \left(
1 +S
\right)
- \frac{5}{3} -\frac{5r}{6},
\end{equation}
where we have defined
\beq
S=\frac{\sigma_{\rm osc} \sigma_{\rm osc}^{\prime\prime}}{\sigma_{\rm osc}^{\prime 2}}
\eeq
with $S=0$ for a purely quadratic potential.
Also notice that, although the curvaton scenario generally generates
large non-gaussianity with $f_{\rm NL} \gtrsim \mathcal{O}(1)$, the
non-linearity parameter $f_{\rm NL}$ can be very small in the presence
of the non-linear evolution of the curvaton field\cite{kesn,Sasaki:2006kq}, which can render the
term $1 + S \simeq 0$~.

\begin{figure}
\centerline{\resizebox{75mm}{!}{\includegraphics{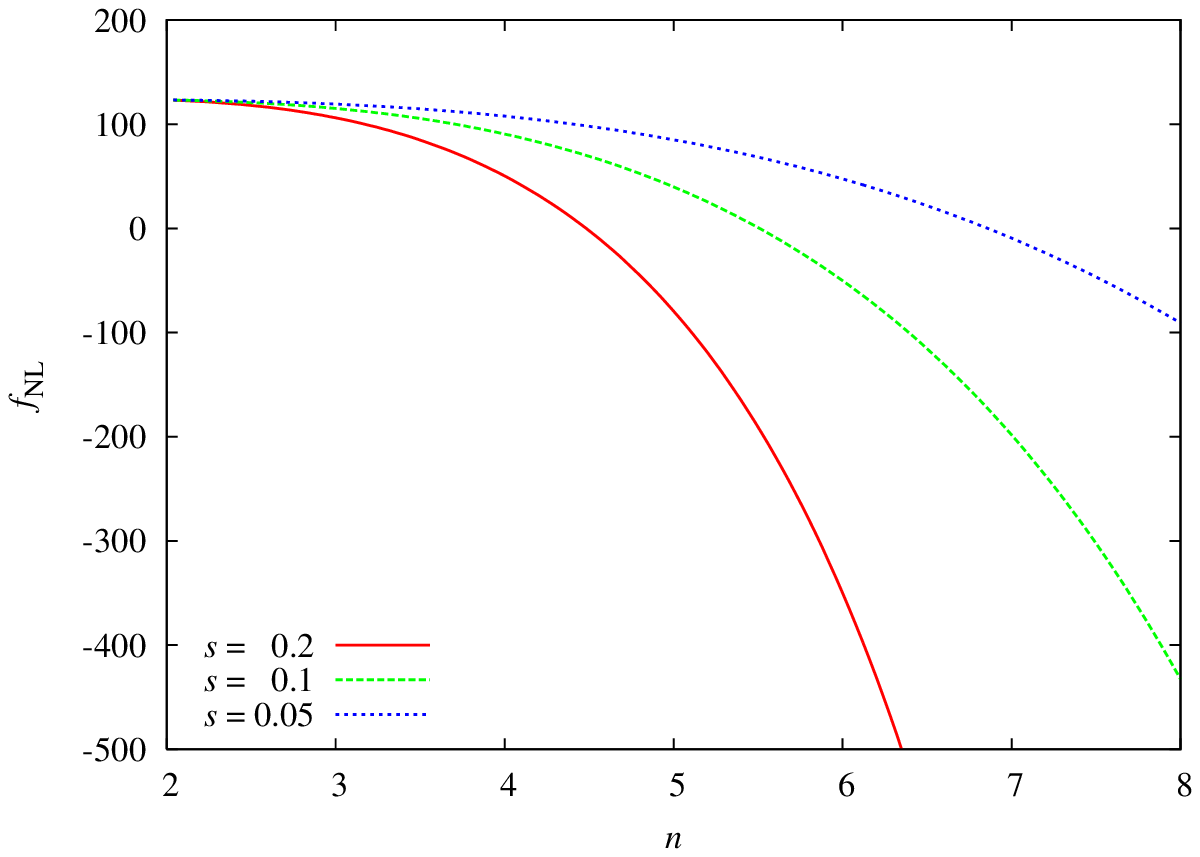}}
\resizebox{75mm}{!}{\includegraphics{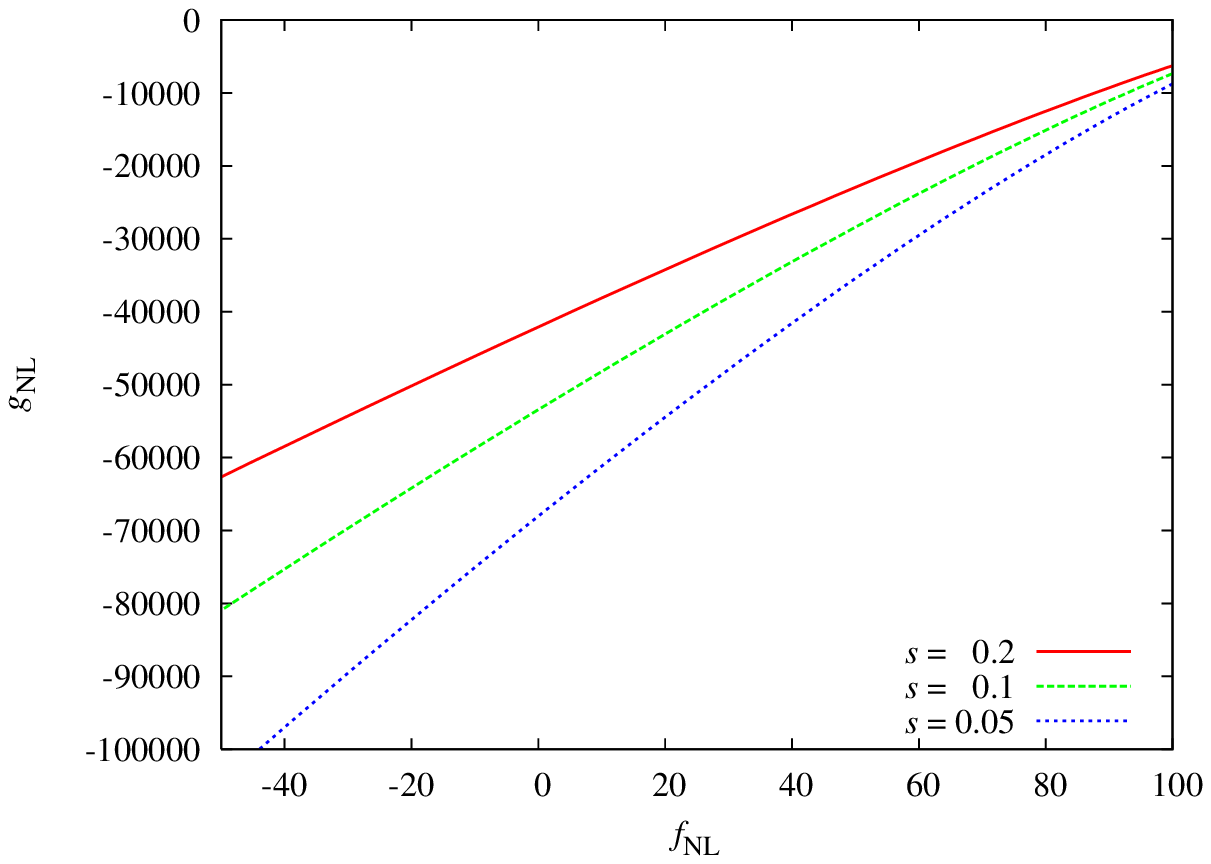}}}
\caption{(Left) Plot of $f_{\rm NL}$ as a function of $n$ for several
  values of $s$.  (Right) Plot of $g_{\rm NL}$ as a function of
  $f_{\rm NL}$ for several values of $s$.  Notice that $f_{\rm NL}$
  and $n$ have one-to-one correspondence.  In both panels, $r=0.01$. }
\label{fig:fNL_gNL_n}
\end{figure}

The non-linearity parameter $g_{\rm NL}$ associated with the trispectrum can be
written as
\beq
\label{eq:gNL}
g_{\rm NL} = \frac{25}{54}
\left[
\frac{9}{4r^2}  \left(
\frac{\sigma_{\rm osc}^2 \sigma_{\rm osc}^{\prime\prime\prime}}
{\sigma_{\rm osc}^{\prime 3}}
+
3S
\right)
-\frac{9}{r}
\left(
1
+
S
\right)
+\frac{1}{2} \left(
1
-
9S
\right)
+10r + 3r^2
\right].
\eeq
As one can easily see, even if the non-linear evolution of $\sigma$
cancels to give a very small $f_{\rm NL}$, such a cancellation does not
necessarily occur for $g_{\rm NL}$. Examples\cite{kett} of the behavior of the
non-linearity parameters for various types of self-ineraction are depicted in Fig.\ref{fig:fNL_gNL_n}.

\section{General results}
\label{numresults}

\begin{figure}
\centerline{\includegraphics[width=8 cm,angle=270]{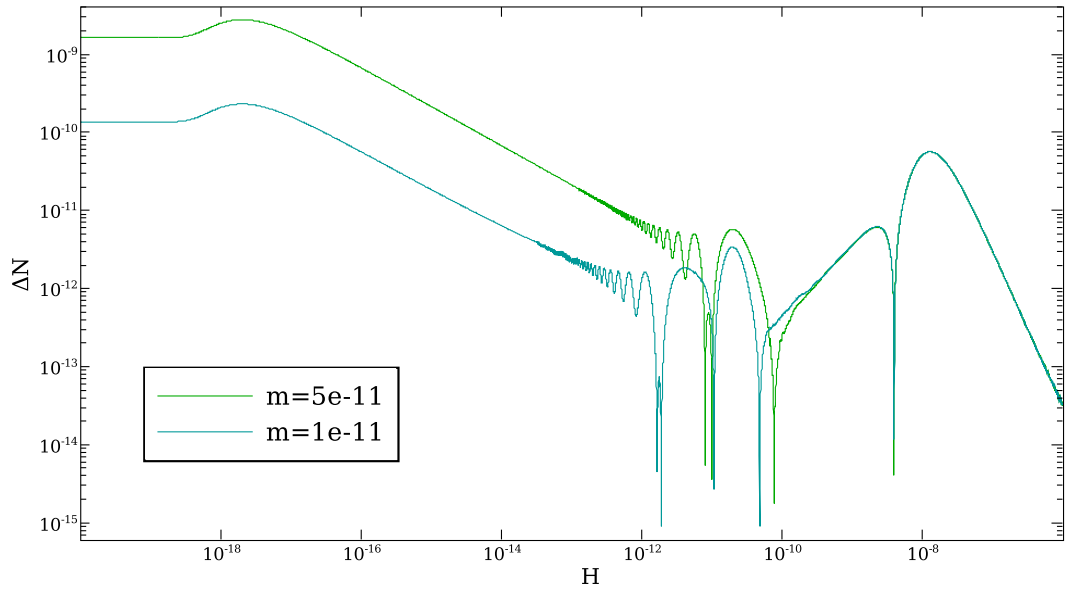}}
\caption{$\Delta N$ as a function of $H$, (time
evolves from right to left), for the masses $m =
10^{-11}$ (lower curve) and $m=5 \cdot 10^{-11}$ (upper curve). In both examples $H_* = 10^{-6}$, $n =
4$, $\Gamma = 10^{-18}$ and $r_*=10^{-10}$.}
\label{fig:1}
\end{figure}

Let us now relax the assumption that the deviations from
the quadratic case are small.
The numerical
solutions of the equations of motion (\ref{frw1}) exhibit complicated
behaviour not qualitatively present in the simplified analytical
approximation. To demonstrate this, in Fig.\ \ref{fig:1} we plot
$\Delta N$ as a function of $H(t)$, or the inverse of time, for fixed initial values of $H_*$, $r_*$ (see (\ref{r_star})) and
$\Gamma$ for two different curvaton mass values. For different
masses the moment of transition from the non-renormalizable part of
the potential to the quadratic part of the potential is different,
and this affects strongly the final value of $\Delta N$ which is
dictated by the duration of oscillations in the non-quadratic
regime.
Deep in the quadratic regime the oscillations become faster and faster
so that $\Delta N$'s evolution is given by a scaling law\cite{us1}. However,
before the quadratic term in the potential starts to dominate, $\Delta N$ shows complicated oscillatory behaviour
that can be only tracked numerically.

From Fig.\ \ref{fig:1} it is clear that as the field value
oscillates in time, so does $\Delta N$. In the non-quadratic regime
$\Delta N$ oscillates with a large amplitude. If the transition to
the quadratic regime is slow compared to the oscillations in the
non-quadratic regime, the transition averages over several
oscillations. As a consequence, the final value of $\Delta N$ will
be a non-oscillatory function of the model parameters. However, if
the oscillation frequency in the non-quadratic potential is slow,
and the transition to the quadratic oscillations is rapid, then the
phase of the non-quadratic oscillation affects the final value of
$\Delta N$. If the parameters happen to be such that the transition
to the quadratic regime occurs at a maximum of the oscillation, a
relatively high value of $\Delta N$ freezes out. Similarly, if the
transition occurs at a minimum of the oscillation cycle, the final
value of $\Delta N$ will be much smaller. If the parameters
governing the moment of transition, such as the curvaton mass $m$,
are changed continuously, then the phase of the non-quadratic
oscillation during the transition also changes continuously. In the
space of the parameters this results in an oscillatory pattern in
$\Delta N$. This behaviour can be understood by observing that the
curvaton energy density at the beginning of its oscillations in the non-quadratic
part of the potential can not be
expressed in terms of an amplitude of the envelope alone but also
depends on the phase of the oscillation, or equivalently on both the field
$\sigma$ and its time derivative $\dot{\sigma}$, in a non-trivial way.
In effect, these act as two independent dynamical degrees of
freedom. If the transition from the interaction dominated part to
the quadratic region takes place at this stage, the initial
variation of the curvaton value $\sigma_{*}$ can therefore translate
in a non-trivial fashion into the final value of the curvature
perturbation.

For a potential with $V\sim
\sigma^{n+4}$ no oscillatory solutions exist if $n\geq6$. This means
that in the non-quadratic regime the curvaton merely decays and
hence no oscillations in $\Delta N$ occur. In \cite{us1,us2} we have
scanned the parameter space $(m,n,\Gamma)$ to find the regions that
yield $\zeta\simeq 10^{-5}$ and acceptably small non-gaussianity while
being consistent with the slow-roll assumption. For fixed parameter values, the result can be mapped
out as a region in the space of the inflation scale $H_*$ and the initial relative curvaton
energy fraction, $r_*$. We also bound
$H_*$ from above by $H_* \lesssim 10^{-5}$, in order to prevent the excessive
production of primordial tensor modes and to keep the inflaton perturbation negligible.

\begin{figure}
\centerline{\includegraphics[height=9cm]{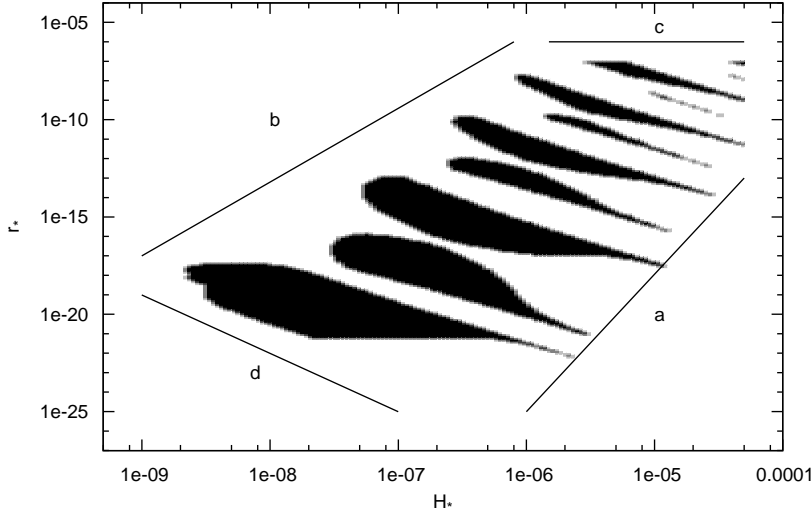}}
\caption{A schematic illustration of the different cuts limiting the allowed area in the parameter space
of the inflation scale $H_*$ and the initial curvaton energy density relative to the total density $r_*$ .}
\label{fig:2}
\end{figure}

The experimental limits for $\fnl$ are given by the WMAP 5-year data \cite{wmap}, $-9 <
\fnl < 111$; we
also require that $-3.5\times10^5 < \gnl <
8.2\times10^5$ as given in \cite{uros}.
The schematic outcome of the scan is depicted in Fig.\ \ref{fig:2}.

The observational
limits for $\fnl$ and $\gnl$ constrain the allowed area in the very
subdominant regions of the parameter space, depicted in Fig.
\ref{fig:2} by the line \emph{a}. Other constraints shown in
Fig.~\ref{fig:2} arise from the internal consistency of the self-interacting curvaton scenario. The bound \emph{b} is obtained because otherwise the
initial perturbations would be too small to produce the observed
amplitude. The bound \emph{c} reflects
the requirement that the curvaton should be massless, or $V'' < H_*^2$,
which is necessary for the generation of curvaton perturbations during
inflation. Because of the subdominance of the curvaton, the realistic
bound should arguably be a few orders of magnitude tighter. However,
a change of an order of magnitude moves the actual cut by a very
small amount in the log-log plots.
Finally, the bound \emph{d}
guarantees the absence of the isocurvature modes in dark matter perturbations and
corresponds to the limit on the curvaton decay width given in
(\ref{eq:isocurvature}).
Detailed figures that show the allowed regions of $H_*$ and $r_*$ for different parameter values can be found in\cite{us2}.

\section{TeV mass curvaton}
\label{tevcurvaton}

It is of particular interest to consider the self-interacting curvaton with a mass $m\simeq 1$ TeV.
Examples of such a curvaton could be found among the MSSM flat directions \cite{mssmflatd} or light moduli fields
of string theories. Let us therefore
fix $m=1$ TeV and demonstrate first that a simple non-interacting or quadratic form does not give rise to a
consistent curvaton model, as discussed in\cite{us3}.

For a quadratic curvaton potential one can write the perturbation as
\beq
 \zeta \sim \frac{H_*}{\sigma_*} r_\mathrm{eff} \simeq 10^{-5}\; ,
\eeq
where $H_*/\sigma_*$ gives the initial perturbation amplitude in the curvaton, and $r_\mathrm{eff}$ is
the efficiency factor that can be approximated quite well by the energy fraction at the curvaton decay \cite{LUW}:
\beq\label{refftev} 
r_\mathrm{eff} \approx \rdec \equiv \left.\frac{\rho_\sigma}{\rhor+\rho_\sigma}\right|_{\hbox{decay}} \; .
\eeq
Relating $\sigma_*$ and $r_*$ from $\frac{1}{2}m^2\sigma_*^2 / 3\Mp^2H_*^2 \simeq r_*$, and noting that $\rdec<1$, we find the constraint on the initial curvaton energy fraction
\begin{equation}
r_* <  \frac{1}{6}\frac{m^2}{\zeta^2 \Mp^2} \, . \label{eq:rzeta}
\end{equation}
In the free curvaton case $\rdec$ also determines non-gaussianity through the simple relation \cite{LUW}
$\fnl = {5}/{4\rdec}$. Very roughly, observationally $|\fnl|<100$, which implies the constraint
\begin{equation}
r_* >  \frac{10^{-4}}{6}\frac{m^2}{\zeta^2 \Mp^2} \, . \label{eq:rnongauss}
\end{equation}

The limits (\ref{eq:rzeta}) and (\ref{eq:rnongauss}) are well known. However, there is more.
Since the observed perturbations are adiabatic to great accuracy, the curvaton must decay before dark matter decouples.
For each set of the initial conditions,
$(H_*, r_*)$, there is a relation between $\rdec$ in (\ref{refftev}) and the effective decay constant $\Gamma$ given by
the fact that decay time is defined as $H=\Gamma$.
Here we assume implicitly a perturbative curvaton decay, but $\Gamma$ could stand for
any effective inverse decay time and thus the following discussion
should hold, at least roughly, also for a non-perturbative curvaton
decay as discussed in \cite{curvatondecres} (note however that non-perturbative curvaton decay
could turn out to be a source of a considerable non-gaussianity \cite{Chambers:2009ki}).

\begin{figure}
\centerline{\includegraphics[width=12cm]{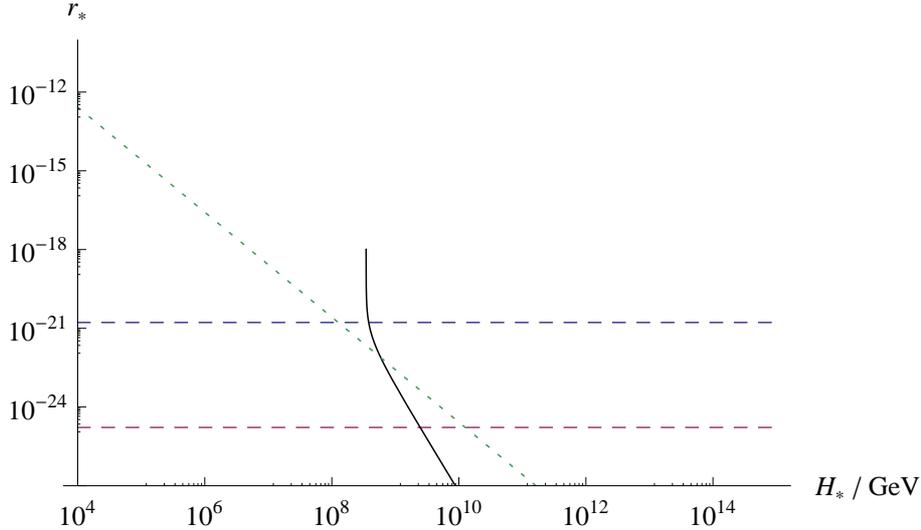}}
\caption{Parameter space of the quadratic curvaton. $r_*$ must be above the lower horizontal dashed line to produce $\zeta \sim 10^{-5}$ (equation (\ref{eq:rzeta})) and below the blueupper horizontal dashed line to produce small enough $\fnl$ (equation (\ref{eq:rnongauss})). Furthermore, $\Gamma$ is constrained from above, and thus only the parameter space to the right of the black solid line is allowed (equation~(\ref{eq:rgamma})). The green dotted line illustrates the equality of the mass term and a possible self-interaction term in the potential (equation (\ref{eq:req})) for $n=4$. For smaller values of $n$ the line moves further to the left. To the right of the dotted line the self-interaction dominates, and thus practically in all of the allowed parameter space the self-interaction must be taken into account.}
\label{fig:3}
\end{figure}

The exact evolution of the energy densities is difficult to solve analytically. However, we can approximate the
curvaton evolution by dividing it up to three phases:
\begin{enumerate}
 \item When $V'' = m^2 < H^2$, the curvaton is effectively massless, so the field value stays constant, $\sigma = \sigma_*$.
 \item When $V'' = m^2 > H^2$, the curvaton oscillates in the quadratic potential, and thus its energy density approximately scales as $\rho_\sigma \propto a^{-3}$.
 \item The curvaton oscillates until $H = \Gamma$, whence it decays.
\end{enumerate}
Solving the Friedmann equation for the regime where $m^2 > H^2$ then yields
\[ \frac{a(H)}{a_*} = \sqrt{\frac{H_*}{H}} \left\{ 1 + \frac{r_*}{4} \left[Â \frac{H_*^2}{m\sqrt{Hm}}-1\right]\right\} + \mathcal{O}\left( r_*^2\right)
\; .\]
Using the above result we can solve for $r_*$ to find
\begin{equation}
r_* = \frac{m\sqrt{m \Gamma}}{H_*^2} \frac{6 \left( \frac{\Mp}{m}\right)^2 \zeta^2}{\frac{H_*^2}{m\sqrt{m\Gamma}} - 12 \left( \frac{\Mp}{m}\right)^2\zeta^2} \; . \label{eq:rgamma}
\end{equation}
We need to check whether, given the constraints discussed above, the
self-interactions can be neglected if $m\simeq 1$ TeV. Thus, adopting the form of the potential given in (\ref{curvatonpot}),
in order for the quadratic assumption to be consistent,
we should require that
\beq
\frac{1}{2}m^2 \sigma^2 \gg \frac{\sigma^{n+4}}{\Mp^n}
\eeq
throughout the evolution. Since the energy density of the quadratic field decreases monoton\-ously, it is sufficient to apply this requirement
only for the initial conditions.
Solving for $r_*$ such that the magnitudes of the quadratic and non-quadratic terms are equal, we find the condition
\begin{equation}
 \label{eq:req}
r_* = \frac{m^2}{3\Mp^2H_*^2} \left( \frac{m^2 \Mp^n}{2} \right)^\frac{2}{n+2} \; .
\end{equation}
We have plotted this condition for $n=4$ in figure \ref{fig:3} as the diagonal dotted line. To the right of it, the non-quadratic term
dominates initially.
As can be seen in figure \ref{fig:3}, there is practically no allowed region in the parameter space where the quadratic assumption
would even approximately apply. For smaller values of $n$,
the self-interaction becomes important even for much smaller values of $H_*$ and $r_*$, and thus, there is no quadratic regime left in the
parameter space.

We thus may conclude that even if the curvaton self-interactions were very weak,
a purely quadratic potential would not be a consistent approximation for a mass $m\simeq 1\,\mathrm{TeV}$;
instead, the effects of the self-interactions need to be taken into account. These change the dynamics of the curvaton in a significant way.
Moreover, as discussed in \cite{us3}, a scan of the parameter space reveals that only $n=4$ potential with $V\sim \sigma^8$ has any
allowed parameter space. In addition, in order to obtain a correct perturbation amplitude,
the decay width $\Gamma$ should be in the range $10^{-15} - 10^{-17}$ GeV. For most particle physics models,
this would be a rather small decay width. We estimate\cite{us3} roughly that in the MSSM, where the non-zero curvaton background
provides masses to other particles and hence gives rise to a kinematical
blocking\cite{AM4}, one could obtain widths of the order $\Gamma\sim 10^{-12}$. However, a detailed and more proper
calculation is required to settle the issue.

\section{Discussion}
\label{jaarittelua}
It may appear surprising that even very small deviations from the quadratic form of the curvaton potential can affect
the curvature perturbation in a significant way. However, one should bear in mind that the small curvature perturbation is really
the difference of two large numbers. The number of e-folds generated during curvaton oscillations is typically $N\sim {\cal O}(10)$,
whereas the difference that gives rise to the non-gaussianity is $\Delta N\lesssim 10^{-8}$. Since
self-interactions imply non-linearities in the
evolution of the curvaton field and in the number of e-folds $N$, one can understand that even small changes can have
profound effects in the difference $\Delta N$. In particular, as discussed here, the non-gaussianities turn out to be quite different as compared with the simplest quadratic model.
There the magnitude of $\fnl$ in the limit $\rdec \ll 1$ is determined by the curvaton energy density at the time of its decay, $\fnl \sim 1/\rdec$.
However, with self-interactions the prediction for $\fnl$ can significantly deviate from this simple estimate.
Even if $\rdec \ll 1$, there exists regions in the parameter space with $|\fnl| < \mathcal{O}(1)$.
This is because the value of $\fnl$ oscillates and changes its sign. Nevertheless,
 $\gnl$ can then be
very large and one has a rather non-trivial non-Gaussian statistics characterized by a large trispectrum and a vanishing bispectrum.
Such a situation, discussed already in \cite{kett}, appears to be rather generic in self-interacting curvaton models, and is possible for a wide, albeit restricted, range of model parameters.
Large non-gaussianities can be generated even if the curvaton dominates the energy density at the time of its decay.
In general, in the presence of self-interactions the relative signs of $\fnl$ and $\gnl$ and
the functional relation between them are typically modified from the quadratic case. Thus the
non-linearity parameters taken together, in possible conjunction of other cosmological observables such as tensor perturbations,
may offer the best prospects for constraining the physical properties of the curvaton.

A TeV mass curvaton is a rather special case.  An important constraint, valid also for higher mass curvatons, is that it has to decay before the CDM freeze-out. This, together with
 observational constraints, fixes the range of the initial conditions for the curvaton field which turn out to be such that the quadratic term in the curvaton potential cannot dominate over possible higher-order terms for the whole dynamical range. One finds\cite{us3} that the only viable curvaton potential that satisfies all the constraints is $V=m^2\sigma^2/2+\sigma^8/M^4$. Moreover, the curvaton decay rate should be in the range $\Gamma=10^{-15}- 10^{-17}$ GeV. Note that in the case where the curvaton energy density is subdominant at the time of decay, the curvaton does not necessarily have to decay before baryogenesis, which can be a process that takes place among the inflaton decay products. However, the decay should be able to produce thermal CDM particles so that the CDM perturbation is adiabatic.

Note also that what really matters is the equation of state, not the time of decay.
Thus if the curvaton decays too early, the perturbations might still generated if the
decay products have the equation of state of matter. An example of this could be the MSSM flat direction fragmenting into Q-balls, which would then slowly decay.

\section*{Acknowledgements}
I should like to thank Sami Nurmi, Gerasimos Rigopoulos, Olli Taanila, and Tomo Takahashi for many enjoyable discussions
on self-interacting curvatons. This work is supported by the Academy of Finland
grants 218322 and 131454.

%

\end{document}